\documentclass[twocolumn,showpacs,prb]{revtex4}
\usepackage{graphicx}%
\usepackage{dcolumn}
\usepackage{amsmath}

\makeatletter
\def\btt#1{\texttt{\@backslashchar#1}}%
\DeclareRobustCommand\bblash{\btt{\@backslashchar}}%
\makeatother

\begin{document}

\title[Short Title]{A phenomenological theory of zero-energy Andreev resonant states}

\author{Yasuhiro Asano}
\email{asano@eng.hokudai.ac.jp}
\affiliation{%
Department of Applied Physics, Hokkaido University, 
Sapporo 060-8628, Japan
}%

\author{Yukio Tanaka}
\affiliation{
Department of Applied Physics, Nagoya University, 
Nagoya 464-8603, Japan}%

\author{Satoshi Kashiwaya}
\affiliation{National Institute of Advanced Industrial Science and Technology, 
Tsukuba, 305-8568, Japan }%

\date{\today}

\begin{abstract}
A conceptual consideration is given to a zero-energy state (ZES) at the surface of 
unconventional superconductors. 
The reflection coefficients in normal-metal / superconductor 
(NS) junctions are calculated based on a phenomenological description 
of the reflection processes of a quasiparticle.
The phenomenological theory reveals the importance of the sign change in 
the pair potential for the formation of the ZES.
The ZES is observed as the zero-bias conductance peak (ZBCP) in the differential 
conductance of NS junctions. The split of the ZBCP due to broken time-reversal symmetry states 
is naturally understood in the present theory. We also discuss effects of external magnetic 
fields on the ZBCP.
\end{abstract}

% PACS, 
\maketitle

\section{introduction}
Transport phenomena in unconventional superconductors have attracted considerable interst 
in recent years because high-$T_c$ superconductors may have the $d$ wave 
pairing symmetry.~\cite{tsuei,sigrist,wollman}
The unconventional pairing symmetry causes the anisotropy 
in transport properties such as the electric conductance and the thermal 
conductivity.~\cite{tanatar,izawa}
In normal-metal / high-$T_c$ superconductor junctions, for instance, 
the shape of the differential conductance reflects the density of states when
the $a$ axis of high-$T_c$ materials is perpendicular to the junction interface.
When $a$ axis deviates from the interface normal, on the other hand,  
the conductance shows a large peak at the zero 
bias-voltage.~\cite{tanaka0,RPP,lofwander,kashiwaya,Alff,Wang,wei,iguchi,geerk,mao,Ekin,Sawa1,Aubin} Such anisotropy in the conductance is now explained by the formation of
a zero-energy state (ZES)~\cite{tanaka0,hu} at the interface of junctions. 
Since the ZES appears just on the Fermi energy, 
it drastically affects transport properties through the interface of 
unconventional superconductor junctions. 
The low-temperature anomaly of the Josephson current between the two 
unconventional superconductors is explained
in terms of the resonant tunneling of Cooper pairs via the 
ZES.~\cite{barash,tanaka1,tanaka2,tanaka3,asano01-3}
So far a considerable number of studies have been made on the ZES itself and 
related phenomena of transport properties in both spin-singlet and spin-triplet 
unconventional superconductor
junctions.~\cite{RPP,lofwander,asano01-2,Buch,asano02-2,Yama1,Yama2,Yama3,zhu,kashi-1,zutic,Y1,Y2,H1,H2,T1,circuit,Kuroki1,Kuroki2,Kusakabe,Honer,St,Sen,asano8,asano9,tanaka4,matsumoto,shirai}

The conductance in normal-metal / superconductor (NS) junctions
is calculated from the normal and the Andreev reflection~\cite{andreev} coefficients which 
are obtained by solving the Bogoliubov-de Gennes (BdG) equation~\cite{degennes} under
appropriate boundary conditions at the junction interface. 
Consequently we easily find the zero-bias conductance peak (ZBCP) in NS junctions of high-$T_c$
superconductors.~\cite{tanaka0} Although the algebra itself is straightforward, it is not easy 
to understand the physics behind the calculation. 
In a previous paper,~\cite{tkc01} we briefly discussed reasons for the appearance of the ZBCP
by a phenomenological argument.
The phenomenological analysis has several advantages.
For instance, it shows the importance of the unconventional 
pairing symmetry for the formation of the ZES without directly solving the BdG equation. 
Moreover we easily understand that the ZES
is a result of the interference effect of a quasiparticle.
The applicability of the analysis in the previous paper, however, 
is very limited because of its simplicity.  

In this paper, we reconstruct the phenomenological theory of the
Andreev reflection to meet the mathematical accuracy. 
We calculate the reflection coefficients of an electronlike
quasiparticle incident from a normal metal into a NS interface.
Near the junction interface, a quasiparticle suffers two kinds of 
reflection: (i) the normal reflection by the barrier potential at the NS interface 
and (ii) the Andreev reflection by the pair potential in the superconductor.
In the present theory, we consider the two reflections separately to calculate the
transport coefficients. 
As a consequence, the Andreev reflection coefficient is decomposed
into a series expansion with respect to the normal reflection probability of NS junctions.
The expression of the Andreev reflection probability enables us to understand the 
importance of the unconventional pairing symmetry for the formation of the ZES.
In unconventional superconductors, the pair potential
in the electron branch ($\Delta_+$) differs from that in 
the hole branch ($\Delta_-$).
The Andreev reflection probability at the zero-energy
is expressed as the summation of the alternating series 
when $\Delta_+$ and $\Delta_-$ have the same sign with each other. 
In this case, the zero-bias conductance becomes
a small value proportional to $|t_N|^4$, where $|t_N|^2$ is the normal transmission 
probability of junctions.
On the other hand when $\Delta_+ \Delta_- <0$, all the expansion 
series have the same sign and the conductance has a large peak at the zero-bias. 
The phenomenological theory can be applied to superconductors with
a broken time-reversal symmetry state (BTRSS)~\cite{fogelstrom,covington,biswas,dagan,sharoni,kohen,matsumoto2,laughlin,kashi95,TJ1,TJ2,Tanuma2001,lubimova,kitaura}
 and NS junctions under external magnetic fields.~\cite{fogelstrom,YT021,YT022,YT023}

This paper is organized as follows. In Sec.~II, the Andreev and the normal
reflection coefficients are derived from a phenomenological description
of a quasiparticle's motion near the NS interface.
In Sec.~III, we discuss the conductance peaks in NS junctions.
A relation between the broken time-reversal symmetry states and the peak 
position in the conductance is discussed in Sec.~IV. we apply the phenomenological 
theory to NS junctions under magnetic fields in Sec.~V. 
In Sec.~VI, we summarize this paper.

\section{Quasiparticle's motion near NS interfaces}
\begin{figure}[htbp]
\begin{center}
\includegraphics[width=8.0cm]{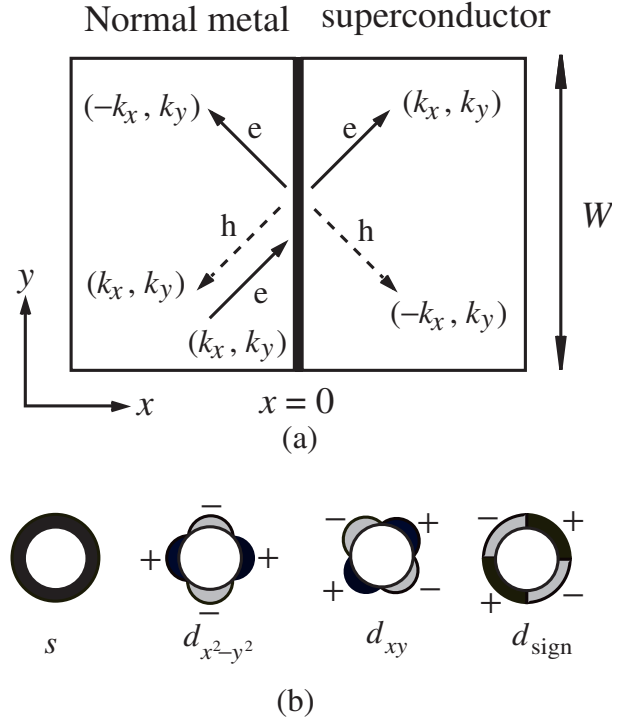}
\end{center}
\caption{
A normal-metal / superconductor junction is illustrated in (a).
The trajectories of a quasiparticle in the electron branch and those in the hole 
branch are denoted by solid and broken lines, respectively.
In (b), the pair potentials of $s$, $d_{x^2-y^2}$ and $d_{xy}$ wave symmetries
are schematically illustrated. 
}
\label{system}
\end{figure}

Let us consider two-dimensional NS junctions as shown in Fig.~\ref{system},
where a normal-metal ($x<0$) and a superconductor ($x>0$) 
are separated by a potential barrier $V(\boldsymbol{r})=V_0 \delta(x)$.
We assume the periodic boundary condition in the $y$ direction and the 
width of the junction is $W$.
The NS junctions are described by the Bogoliubov-de Gennes equation,~\cite{degennes}
\begin{align}
\int d{\boldsymbol{r}'} &\left( \begin{array}{cc}
\delta(\boldsymbol{r}-\boldsymbol{r}')h_0(\boldsymbol{r}') & \Delta(\boldsymbol{r},\boldsymbol{r}')
e^{i\varphi_s} \\
\Delta^\ast(\boldsymbol{r},\boldsymbol{r}')e^{-i\varphi_s} 
& -\delta(\boldsymbol{r}-\boldsymbol{r}')
h_0(\boldsymbol{r}')\end{array}\right) \nonumber \\
& \times\left( \begin{array}{c} u(\boldsymbol{r}') \\  v(\boldsymbol{r}') \end{array} \right) 
 = E  \left( \begin{array}{c} u(\boldsymbol{r}) \\  v(\boldsymbol{r}) \end{array} \right),
\end{align}
\begin{align}
h_0(\boldsymbol{r}) =& -\frac{\hbar^2 \nabla^2}{2m} + V_b\delta(x) - \mu_F, \\
\Delta(\boldsymbol{R_c},\boldsymbol{r}_r)= &\left\{\begin{array}{ccc}  
\frac{1}{V_{vol}}\sum_{\boldsymbol{k}} \Delta(\boldsymbol{k}) e^{i\boldsymbol{k}\cdot 
\boldsymbol{r}_r} & : &
 X_c > 0\\
0  & : &  X_c < 0 
\end{array}\right.,
\end{align}
where $\varphi_s$ is a macroscopic phase of the superconductor, 
$\boldsymbol{R}_c=(X_c,Y_c)=(\boldsymbol{r}+\boldsymbol{r}')/2$ and 
$\boldsymbol{r}_r=\boldsymbol{r}-\boldsymbol{r}'$.  
Here we assume spin-singlet superconductors for simplicity.
The argument in the following can be extended to spin-triplet superconductors as 
shown in Appendix.
When an electronlike quasiparticle 
is incident from the normal metal as shown in Fig.~\ref{system}, 
the wave function in the normal metal is given by,
\begin{align}
\Psi_N(\boldsymbol{r}) =& \left[\left( \begin{array}{c} 1 \\ 0 \end{array}\right) e^{ik_xx} 
+\left( \begin{array}{c} 1 \\ 0 \end{array}\right) e^{-ik_xx} r^{ee}\right.
\nonumber\\
+& \left. \left( \begin{array}{c} 0 \\ 1 \end{array}\right) e^{ik_xx} r^{he}
\right] \frac{e^{ik_yy}}{\sqrt{W}}. \label{wfn}
\end{align}
where $k_x$ and $k_y$ are the wave numbers on the Fermi surface and they satisfy 
$k_x^2+k_y^2=k_F^2$ with $k_F$ being the Fermi wave number.
Throughout this paper we assume that $E \sim \Delta_0 \ll \mu_F$, where $\Delta_0$ is the 
amplitude of the pair potential and $E$ is the energy of a quasiparticle measured from 
the Fermi energy, $\mu_F=\hbar^2k_F^2/(2m)$.
In Eq.~(\ref{wfn}), $r^{ee}$ and $r^{he}$ are the normal and the Andreev reflection coefficients,
respectively.  

When a quasiparticle is incident from the normal metal in the electron branch, 
directions of the outgoing waves are indicated by arrows as shown in Fig.~\ref{system}. 
The trajectories of a quasiparticle in the electron branch and those in the hole 
branches are denoted by solid and broken lines, respectively. In the normal metal,
a velocity component perpendicular to the interface changes
its sign in the normal reflection, whereas all velocity components change signs in the 
Andreev reflection.
In the superconductor, the wave number in the electron branch is $(k_x,k_y)$, but
that in the hole branch becomes $(-k_x, k_y)$.
In unconventional superconductors, the
pair potential in the electron branch ($\Delta_+\equiv \Delta(k_x,k_y)$) differs from 
that in the hole branch ($\Delta_-\equiv \Delta(-k_x,k_y)$).
Therefore the wave function in the superconductor is described by these two pair potentials,
\begin{align}
\Psi_S(\boldsymbol{r}) =& 
\left[ \left(\begin{array}{c}  u_+ \\ e^{-i\phi_+}e^{-i\varphi_s} v_+ \end{array} \right)
\textrm{e}^{ik^e x} t^{ee} \right.\nonumber \\+&\left. 
\left(\begin{array}{c} e^{i\phi_-}e^{i\varphi_s} v_- \\  u_- \end{array} \right)
\textrm{e}^{-ik^h x} t^{he} \right] 
\frac{e^{ik_y y}}{\sqrt{W}}, \label{wfs}\\
u_\pm (v_\pm) =& \sqrt{ \frac{1}{2} \left( 1+ (-)\frac{\Omega_\pm}{E}\right)},\\
\textrm{e}^{i\phi_\pm}\equiv &\frac{\Delta_\pm}{|\Delta_\pm|},\label{s-p}\\
k^{e(h)} = & \left[ k_x^2 +(-) k_F^2 \frac{\sqrt{E^2-|\Delta_{+(-)}|^2}}{\mu_F}\right]^{1/2},\\
\Omega_\pm =& \sqrt{ E^2 - |\Delta_\pm|^2}, 
\end{align}
where $t^{ee}$ ($t^{he}$) is the transmission coefficient to the electron (hole) branch
in superconductors.
The wave numbers of a quasiparticle are approximately given by 
$k^{e(h)}\approx k_x +(-)i/(2\xi_0)$ for $E \sim 0$, where 
$\xi_0 = \hbar v_F/(\pi \Delta_0)$
is the coherence length and $v_F=\hbar k_F/m$ is the Fermi velocity. 
Thus a quasiparticle penetrates into the superconductor within a range of $\xi_0$.
In Eqs.~(\ref{wfs})-(\ref{s-p}), a phase 
$e^{i\phi_\pm}$ represents the sign (internal phase) of the pair potential and
appears in the wave function in addition to a macroscopic phase of the 
superconductor. 
The transmission and the reflection coefficients
are obtained from the boundary conditions of these wave functions.
Near the junction interface, an incident quasiparticle suffers two kinds of 
reflection: (i) the normal reflection by the barrier potential at the NS interface 
and (ii) the Andreev reflection by the pair potential in the superconductor.
In this paper, we consider separately contributions of the two reflection processes to the 
reflection coefficients. 

We first consider NS junctions with no barrier potential at the interface,
\begin{equation}
z_0 \equiv \frac{V_0}{\hbar v_F}=0,
\end{equation}
where $z_0$ represents the strength of the potential barrier.
The Andreev reflection coefficients become 
\begin{align} 
r^{he}_0 =& -i \nu_+ e^{-i\phi_+}e^{-i\varphi_s},\label{rhe0}\\
r^{eh}_0 =& -i \nu_- e^{i\phi_-}e^{i\varphi_s},\label{reh0}\\
\nu_\pm=& i \frac{ E - \Omega_\pm}{|\Delta_\pm|},\label{nu-def1}
\end{align}
where $r^{he}_0$ is the Andreev reflection coefficients from the electron  
branch to the hole branch in the absence of the potential barrier. 
We also give the Andreev reflection coefficient from the hole branch
to the electron branch ($r^{eh}_0$). 
In the case of $E^2-\Delta_\pm^2 <0$, $\nu_\pm$ can be described as
\begin{align}
\nu_\pm =& \frac{\sqrt{\Delta_\pm^2-E^2}}{|\Delta_\pm|}+i\frac{E}{|\Delta_\pm|},\\
\equiv & \cos\theta_\pm + i \sin\theta_\pm = e^{i\theta_\pm}. \label{nu-def2}
\end{align}
Thus the Andreev reflection coefficients include only the phase information 
in the limit of $z_0=0$.

\begin{figure}[htbp]
\begin{center}
\includegraphics[width=8.0cm]{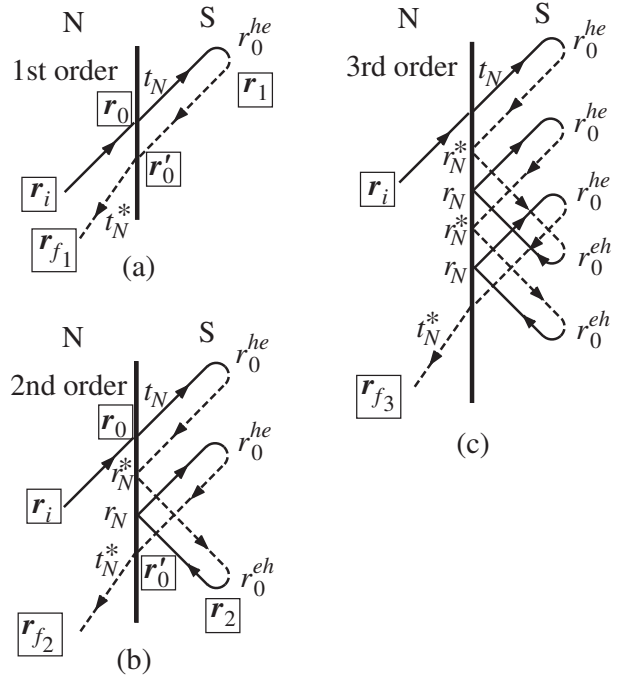}
\end{center}
\caption{
The Andreev reflection processes are decomposed into a series of the
reflections by the pair potential and the barrier potential.
}
\label{process}
\end{figure}

We next consider the reflection by the potential 
barrier in a phenomenological way.
In the presence of the potential barrier, the Andreev reflection processes are 
shown in Fig.~\ref{process}. 
In the electron branch, the normal transmission and the normal reflection coefficients 
of the barrier are calculated to be $t_N=\bar{k}_x/(\bar{k}_x+iz_0)$ and 
$r_N=-iz_0/(\bar{k_x}+iz_0)$, respectively with $\bar{k}_x=k_x/k_F$.
Those in the hole branch are $t_N^\ast$ and $r_N^\ast$.
The Andreev reflection coefficient in the 1st order process is given by,
\begin{equation}
r^{he}(1) = t_N^\ast \cdot r^{he}_0 \cdot t_N.
\end{equation}
At first an electronlike quasiparticle starting from $\boldsymbol{r}_i$ transmits 
into the superconductor through $\boldsymbol{r}_0$, ($t_N$).
In Fig.~\ref{process}, the vectors in real space are surrounded by squares to avoid confusion.
While traveling the superconductor within the range of $\xi_0$, the quasiparticle is 
reflected into the hole branch by the pair potential at $\boldsymbol{r}_1$ ($r^{he}_0$). 
Then the quasiparticle goes back
to the normal metal in the hole branch through $\boldsymbol{r}'_0$ ($t_N^\ast$). 
The 2nd order Andreev reflection process in Fig.~\ref{process} (b) 
can be estimated in the same way,
\begin{align}
r^{he}(2) =& t_N^\ast \cdot A_S \cdot r^{he}_0 \cdot t_N, \\
A_S=& r^{he}_0 \cdot r_N \cdot r^{eh}_0 \cdot r_N^\ast, \\
=& -|r_N|^2 \nu_+\nu_-e^{i(\phi_--\phi_+)}. \label{a-def} 
\end{align}
After the first Andreev reflection into the hole branch, the quasiparticle suffers
the normal reflection ($r_N^\ast$). Next the holelike quasiparticle experiences the 
2nd Andreev reflection 
to the electron branch at $\boldsymbol{r}_2$ ($r^{eh}_0$). 
Then the electronlike quasiparticle suffers the normal reflection ($r_N$) 
followed by the 3rd Andreev reflection into the hole branch ($r^{he}_0$). 
Finally the holelike quasiparticle goes back to the normal metal 
through $\boldsymbol{r}'_0$ ($t_N^\ast$).
We only show the expression of the Andreev reflection coefficient in the 3rd order process,
\begin{align}
r^{he}(3) = t_N^\ast \cdot A_S^2 \cdot r^{he}_0 \cdot t_N. 
\end{align}
The corresponding trajectory is shown in Fig.~\ref{process} (c).
The total Andreev reflection coefficient is obtained by the summation of
these reflection processes up to the infinite order,
\begin{align}
r^{he} =& |t_N|^2 \cdot r^{he}_0 \cdot \sum_{n=1}^\infty  A_S^{n-1}.  \label{rhe1}
%=& \frac{ |t_N|^2 r^{he}_0}{1 - |r_N|^2r^{he}_0r^{eh}_0}. \label{rhe2}
\end{align}
In the similar way, the normal reflection coefficient results in
\begin{align}
r^{ee} =& r_N + t_N^2 \cdot r_N^\ast \cdot r^{he}_0 r^{eh}_0
\sum_{n=1}^{\infty}  A_S^{n-1}.\label{ree1}
\end{align}
%The corresponding trajectories for the normal reflection are shown in Fig.~\ref{pron}.
Although the reflection coefficients in Eqs.~(\ref{rhe1}) and (\ref{ree1}) 
are obtained based on the phenomenological description of a quasiparticle's motion, 
they are mathematically identical to the exact expressions
calculated from the boundary conditions
of the wave functions in the presence of the potential barrier.~\cite{tanaka0}

\section{conductance}
The differential conductance is calculated from the normal and the Andreev 
reflection coefficients,~\cite{blonder,takane}
\begin{equation}
G_{NS}=\left. \frac{2e^2}{h} \sum_{k_y}\left[ 1 - |r^{ee}|^2 + |r^{he}|^2 \right] \right|_{E=eV_{bias}},
\label{btk1}
\end{equation}
where $V_{bias}$ is the bias voltage applied to NS junctions.
We focus on the limit of $E \to 0$ for a while, where
the Andreev reflection probability dominates the zero-bias conductance
because the conductance can be described by
\begin{equation}
G_{NS}=\left. \frac{4e^2}{h} \sum_{k_y} |r^{he}|^2 \right|_{E=eV_{bias}}.
\label{btk2}
\end{equation}
A quasiparticle after the Andreev reflection 
traces back the original trajectory of a quasiparticle before 
the Andreev reflection. 
This is called the retro property
of a quasiparticle.
When we estimate the reflection coefficients in Eq.~(\ref{rhe1}) and(\ref{ree1}),
we only consider the phase factor of the Andreev reflection.
 A quasiparticle, however, may suffer additional phase shift while
moving around the NS interface.
Actually, an electron acquires a phase $e^{i\boldsymbol{k}\cdot (\boldsymbol{r}_1
-\boldsymbol{r}_0)}$ while traveling from $\boldsymbol{r}_0$ to $\boldsymbol{r}_1$
as shown in Fig.~\ref{process} (a).
In addition to this, 
a phase factor $e^{i\boldsymbol{k}\cdot (\boldsymbol{r}'_{0}
-\boldsymbol{r}_{1})}$ is multiplied while traveling from 
$\boldsymbol{r}_1$ to $\boldsymbol{r}'_{0}$ in the hole branch.
These two phase factors exactly cancel each
other out when the retro property holds because $\boldsymbol{r}_{0}=\boldsymbol{r}'_{0}$.
Thus $\boldsymbol{r}_{f_n}$ indicates the same position for all $n$.
In particular for $E=0$, a relation $\boldsymbol{r}_{i}=\boldsymbol{r}_{f_n}$ for all $n$ 
holds, which means the retro property of a quasiparticle in the normal metal.
In the limit of $E \to 0$, we find in Eq.~(\ref{nu-def2}) that $\nu_\pm \to 1$ irrespective
of symmetries of the pair potential.
The Andreev reflection probability becomes
\begin{align}
|r^{he}|^2 =& { |t_N|^4 }\left| \sum_{n=0}^\infty
|r_N|^{2n} \left[-e^{i(\phi_--\phi_+)}\right]^n\right|^2. \label{rhe3}
%\\
%=&\frac{|t_N|^4}{|1 + |r_N|^2 e^{i(\phi_--\phi_+)}|^2}.
\end{align}
Firstly, we consider superconductors where the pair potentials in the two branches 
($\Delta_+$ and $\Delta_-$) have the same sign, (i.e., $e^{i(\phi_--\phi_+)}=1$).
For examples, the pair potentials below satisfy the condition irrespective of the 
wave numbers of a quasiparticle,
\begin{align}
\Delta_{s}(\boldsymbol{k})=&  \Delta_0 & (s\; \text{wave}), \label{s}\\
\Delta_{d_{x^2-y^2}}(\boldsymbol{k})=&  \Delta_0 (\bar{k}_x^2-\bar{k}_y^2) & (d_{x^2-y^2}\; \text{wave}), 
\label{d1}
\end{align}
where $\bar{k}_x ={k}_x /k_F$ and 
$\bar{k}_y ={k}_y/k_F$ 
are the normalized wave number on the Fermi surface in the $x$ and $y$ 
directions, respectively.
The schematic figures of the pair potentials are shown in Fig.~\ref{system} (b).
Equation~(\ref{s}) represents the pair potential of $s$ wave superconductors.
The pair potential in Eq.~(\ref{d1}) is realized in a junction where the $a$ axis of a high-$T_c$
superconductor is perpendicular to the interface normal.
When $e^{i(\phi_--\phi_+)}=1$ is satisfied, 
Eq.~(\ref{rhe3}) becomes the summation of the alternating series.
The Andreev reflection probability results in
\begin{align}
|r^{he}|^2 = \frac{2 |t_N|^4}{(2-|t_N|^2)^2}.\label{arp1}
\end{align}
In low transparent junctions, (i.e., $z_0^2 \gg 1$),
the Andreev reflection probability becomes a small value $|t_N|^4/2 \propto 1/z_0^4$.
Therefore the zero-bias conductance in Eq.~(\ref{btk2}) is proportional to $1/z_0^4$.
Secondly we consider that the signs of the two pair potential are opposite 
to each other. The pair potential
\begin{equation}
\Delta_{d_{xy}}(\boldsymbol{k}) = 2 \Delta_0 \bar{k}_x \bar{k}_y, 
\qquad (d_{xy}\; \textrm{wave}), \label{d2}
\end{equation}
satisfies $e^{i(\phi_--\phi_+)}=-1$ for all wave numbers and is realized in a junction where
the $a$ axis of a high-$T_c$ superconductor is oriented by 45 degrees from
the interface normal.
All the expansion series in Eq.~(\ref{rhe3}) have the same sign and
the Andreev reflection probability becomes
\begin{equation}
|r^{he}|^2 = 1. \label{arp2}
\end{equation}
Thus the zero-bias conductance in Eq.~(\ref{btk2}) takes its maximum value.
The sign of the pair potentials
characterizes the interference effect of a quasiparticle near the NS interface.
For $e^{i(\phi_--\phi_+)}=1$, the alternating series in Eq.~(\ref{rhe3}) reflect 
the destructive interference among the partial waves of a quasiparticle in the expansion series.
Hence the conductance becomes small at the zero-bias. 
On the other hand for $e^{i(\phi_--\phi_+)}=-1$,
the expansion series with the same sign imply that 
the partial waves in the expansion series interfere constructively,  
which leads to the large zero-bias conductance.  
The constructive interference at the interface causes a resonant state 
which is now referred to as the ZES. The Andreev reflection probability 
is unity independent of the normal transmission probability of junctions 
as shown in Eq.~(\ref{arp2}).
This can be interpreted as a result of the resonant transmission of a quasiparticle 
through the ZES.
A microscopic calculation shows that the ZES has a large local density of states 
around $x=\xi_0$ at the zero-energy.~\cite{single}
Similar arguments have been done in normal-metal/insulator/normal-metal/insulator/superconductor
junctions~\cite{belogolovskii} and at the surface of high-$T_c$ superconductors.~\cite{kashi95-2}

In Eq.~(\ref{arp2}), we can explain a large conductance at the zero-bias.
In what follows, we will show that the conductance has a peak structure around the zero-bias.
When $E\neq 0$ but still $E \lesssim \Delta_0$, the degree of resonance is suppressed 
because $\nu_\pm$ is no longer unity as shown in Eq.~(\ref{nu-def2}).
In the superconductor, the argument of the phase cancellation in the round-trip 
between $\boldsymbol{r}_0$ and $\boldsymbol{r}_1$ in Fig.~\ref{process} (a) 
is still valid as far as $E^2-|\Delta_\pm|^2<0$ being satisfied. 
In the electron branch on the way to $\boldsymbol{r}_1$, 
the $x$ component of the wavenumber is given by
\begin{equation}
k^e \simeq k_x +i \frac{k_F}{\bar{k}_x}\frac{\sqrt{|\Delta_+|^2-E^2}}{2\mu_F},
\end{equation}
The real part determines the direction of the quasiparticle's motion.
The inverse of the imaginary part characterizes the dumping  
of the wave function and is roughly estimated to be $\xi_0$. 
It is also shown that $k_x$ is the real part of the wave number in the 
hole branch on the way back to $\boldsymbol{r}_0$. 
The Andreev reflection probability for finite $E$ is given by
\begin{align}
|r^{he}|^2 = 
\frac{|t_N|^4}{|t_N|^4 + 2 |r_N|^2 \left[1+\textrm{Re}\, \nu_+\nu_- e^{i(\phi_--\phi_+)}\right] }.
\end{align}

\begin{figure}[htbp]
\begin{center}
\includegraphics[width=8.0cm]{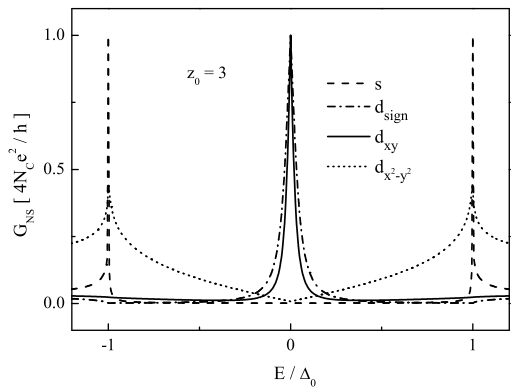}
\end{center}
\caption{
The conductance is plotted as a function of $E$, where $z_0=3$.
The anisotropy of the pair potential in $d_{xy}$ symmetry is taken into account
only through the phase factor $e^{i(\phi_--\phi_+)}$ and
$\boldsymbol{k}$ dependence of the pair potential is 
neglected in the dash-dotted line.
The $d_{xy}$ symmetry is fully taken into account in the solid line.
The conductance for $d_{x^2-y^2}$ symmetry is amplified by 5 times in the dotted line.
}
\label{fig:gns1}
\end{figure}

To make clear a relation between the peak positions of the conductance
and the relative sign of the pair potentials, we consider the pair potential
\begin{equation}
\Delta_{d_{\textrm{sign}}}(\boldsymbol{k}) = \Delta_0 \;\textrm{sgn}(k_xk_y),\label{d3}
\end{equation}
instead of Eq.~(\ref{d2}). 
Here the anisotropy of pair potential is taken into account only through 
the phase $e^{i\phi_\pm}$ and the $\boldsymbol{k}$
dependence of the pair potential is neglected. 
The pair potential in Eq.~(\ref{d3}) is illustrated in Fig.~\ref{system} (b).
We will check the validity of Eq.~(\ref{d3}) later.
The Andreev reflection probability for $\Delta_{d_{\textrm{sign}}}$ becomes
\begin{align}
|r^{he}|^2  
=&\frac{|t_N|^4}{|t_N|^4 + 4 |r_N|^2 \sin^2\theta} = \frac{ E_0^2}{ E^2 + E_0^2},\label{rhe4}\\
E_{0} =& \frac{\Delta_0 |t_N|^2}{2 |r_N|}.\label{pwidth}
\end{align}
where we use a relation $\theta = \theta_+ = \theta_- $ in Eq.~(\ref{nu-def2}). 
The Andreev reflection probability has a peak structure at $E=0$ and
the width of the peak is characterized by $E_0$ which is $\Delta_0/z_0^2$
in the limit of $z_0^2 \gg 1$.
On the other hand in $s$ wave junctions (i.e., $e^{i(\phi_--\phi_+)}=1$), 
we find
\begin{align}
|r^{he}|^2 =& \frac{|t_N|^4}{|t_N|^4 + 4 |r_N|^2 \cos^2\theta}
= \frac{ E_0^2}{ (\Delta_0^2-E^2) + E_0^2}.\label{rhe5}
\end{align}
The Andreev reflection probability has a peak at $E=\Delta_0$ 
reflecting a peak of the bulk density of states in $s$ wave superconductors.
%Above analysis implies that 
%the peak position is determined by $\sin\theta=0$ for $e^{i(\phi_--\phi_+)}=-1$
%and $\cos\theta=0$ for $e^{i(\phi_--\phi_+)}=1$.
In Fig.~\ref{fig:gns1}, we plot the conductance,
where $z_0=3$ and $N_c=Wk_F/\pi$ is the number of propagating channels on the Fermi surface.
The results for $s$ wave junction are indicated by the broken line.
The conductance for $d_{x^2-y^2}$ symmetry in the dotted line is amplified by
5 times.
These conductance has a peak at $E=\Delta_0$ reflecting the bulk density
of states.
The results for $\Delta_{d_{\textrm{sign}}}$ and $d_{xy}$ are shown with 
the dash-dotted line and the solid line, respectively.
There is no significant difference between the conductance for $d_{xy}$ 
symmetry and that for $d_{\textrm{sign}}$ 
 because the relative sign of the two pair potential 
($e^{i(\phi_--\phi_+)}=-1$) dominates the subgap conductance structure. 
Throughout this paper, we describe the pair potential by using the step function at the NS 
interface and neglect its spatial dependence in superconductors.
In real NS junctions, the pair potential is suppressed at the interface 
in the presence of the ZES.~\cite{matsumoto2,Tanuma2001} 
The conductance shape around the zero-bias,
however, almost remains unchanged even if the spatial dependence of the pair potential
is taken into account.~\cite{asai00} 
This is also because relative sign of the two pair potentials 
determines the conductance around the zero-bias.
The spatial dependence of the pair potential may affect the width of the ZBCP through 
$E_0$ in Eq.~(\ref{pwidth}).

We note that there is no remarkable 
differences between the mathematical origin of the peaks at $E=0$ for $e^{i(\phi_--\phi_+)}=-1$ 
and that at $E=\Delta_0$ for $e^{i(\phi_--\phi_+)}=1$.
Actually it is easy to confirm at $E=\Delta_0$ that the Andreev reflection
probability in $s$ wave junctions becomes 
\begin{align}
|r^{he}|^2 =& { |t_N|^4 }\left| \sum_{n=0}^\infty
|r_N|^{2n} \left[e^{i(\phi_--\phi_+)}\right]^n\right|^2. \label{rhe6}
\end{align}
All the expansion series have the same sign for $e^{i(\phi_--\phi_+)}=1$.

In above arguments, we have assumed that the junctions have non zero transmission probabilities.
In the end of this section, we briefly mention that the ZES becomes a real bound state in the limit
of $z_0\to\infty$. A quasiparticle motion is spatially limited 
at the surface of the semi-finite superconductor because of the perfect normal reflection 
by the surface and the Andreev reflection by the pair potential. 
The ZES becomes a bound state because there is no quasiparticle 
excitations which extend into the bulk superconductors at $E=0$. 
In the density of states, such ZES is found as the $\delta$-function peak. 
For finite transmission probability of junctions, the finite propagation 
into normal metals gives a finite life time of the ZES which is given by $\hbar/E_0$. 
On the other hand for $e^{i(\phi_--\phi_+)}=1$, the resonant state at $E=\Delta_0$ 
does not become a bound state because there are excitations
extend into the bulk superconductors at $E=\Delta_0$.
In superconductor/insulator/superconductor (SIS) junctions, the ZES is also a bound state 
irrespective of the transmission probability of junctions. 
The description of the Andreev bound states in SIS junctions was given, for example, 
in Ref.~\onlinecite{barash97}.

\section{pairing without time-reversal symmetry}
\begin{figure}[htbp]
\begin{center}
\includegraphics[width=8.0cm]{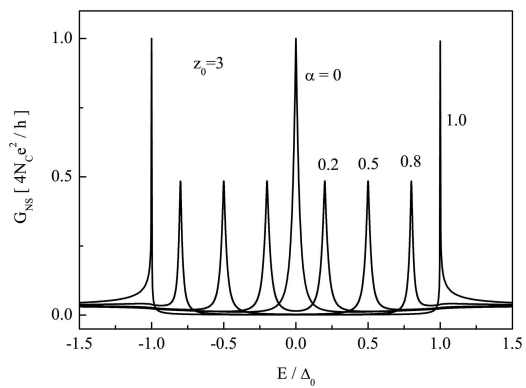}
\end{center}
\caption{
The conductance is plotted as a function of $E$ for $s+id_{xy}$ symmetry, where $z_0=3$.
}
\label{fig:sd}
\end{figure}

In recent experiments, a possibility of
the broken time reversal symmetry state (BTRSS) at the surface of high-$T_c$
superconductors has been discussed.~\cite{covington,biswas,dagan,sharoni,kohen}
These experiments found the split of the ZBCP in the zero magnetic field.
It is pointed out that such surface states may have $s+id_{xy}$~\cite{matsumoto2} or 
$d_{xy}+id_{x^2-y^2}$~\cite{laughlin} pairing symmetry.
Theoretical studies showed the split of the surface density of 
states~\cite{fogelstrom,kashi95,TJ1,TJ2,Tanuma2001,lubimova}
 when $s+id_{xy}$
wave pairing symmetry is assumed at the surface of the $d_{xy}$ wave superconductor. 
Within the present phenomenological 
theory, it is also possible to discuss the split of the conductance peak by the BTRSS
in terms of the shift of the resonance energy.
 We assume the pair potential as
\begin{equation}
\Delta_{s+id_{xy}}(\boldsymbol{k})= \alpha \Delta_0+ i \beta \Delta_{d_{xy}}(\boldsymbol{k}), 
\;\; (s+id_{xy}\;\textrm{wave})\label{sd1},
\end{equation}
with $\alpha^2 + \beta^2=1$.
We find 
\begin{align}
|\Delta_\pm|=&|\Delta|=
\sqrt{\alpha^2\Delta_0^2+\beta^2 \Delta_{d_{xy}}^2(\boldsymbol{k})},\\
e^{i(\phi_--\phi_+)}=&e^{2i\phi_-}= 
\left[\frac{\alpha\Delta_0 -i\beta \Delta_{d_{xy}}}
{|\Delta|}
\right]^2.
\end{align}
In Fig~\ref{fig:sd}, we show the conductance in the $s+id_{xy}$ symmetry
for several $\alpha$.
For $\alpha=0$, the results are identical to the conductance of 
$d_{xy}$ wave junctions in Fig.~\ref{fig:gns1}.
The ZBCP splits into two peaks for $\alpha\neq 0$. The splitting width increases
almost linearly with increasing $\alpha$.
In the limit of $\alpha=1$, the results coincide with the conductance of 
$s$ wave junctions in Fig.~\ref{fig:gns1}.
The peak position can be explained by the expression of the Andreev reflection 
probability
\begin{align}
|r^{he}|^2=& \frac{|t_N|^4}{|t_N|^4 + 4 |r_N|^2\cos^2(\theta+\phi_-)},\label{arp-sd}\\
\cos(\theta+\phi_-) =& \frac{\sqrt{|\Delta|^2-E^2} }{|\Delta|^2} \alpha \Delta_0
 + \frac{E }{|\Delta|^2} \beta \Delta_{d_{xy}},\label{sdcos0}\\ 
\approx &\frac{\sqrt{\Delta_0^2 -E^2}}{\Delta_0} \alpha + \frac{E}{\Delta_0} \beta
\; \textrm{sgn}(k_xk_y). \label{sdcos} 
\end{align}
In the last equation, we replace $\Delta_{d_{xy}}$ by $\Delta_{d_{\textrm{sign}}}$.
The conductance peak (the resonance energy) is expected at an energy which satisfies 
$\cos(\theta+\phi_-)=0$ as shown in Eq.~(\ref{arp-sd}).
The resonance energies for $\alpha=0$ and $\alpha=1$ 
are $E=\Delta_0$ and $E=0$, respectively. 
These resonance energies are independent of the wave numbers.
Consequently the peak heights for $\alpha=0$ and $\alpha=1$ become unity.
The peak heights for finite $\alpha$, however, are always less than unity as shown 
in Fig.~\ref{fig:sd} because the resonance energy depends on wave numbers as shown 
in Eq.~(\ref{sdcos0}).

The positions of the conductance peaks are roughly given by $E=\pm\alpha \Delta_0$,
which can be understood by the resonance condition of $\cos(\theta+\phi_-)=0$ 
in Eq.~(\ref{sdcos}). 
Since the peak position is determined by $\alpha$, relative amplitudes of $s$ and $d_{xy}$
components can be estimated from the peak splitting width observed in experiment.
In the phenomenological theory, effects of the BTRSS on the conductance can be understood
in terms of the shift of the resonance energy.

In theoretical studies, it is shown that the $s+id_{xy}$ wave BTRSS 
splits zero-energy peak of the local density of 
states~\cite{fogelstrom,kashi95,TJ1,TJ2,Tanuma2001,lubimova} and the ZBCP.~\cite{kitaura}
Experimental results are, however, still controversial.
Some experiments reported the split of the ZBCP at the 
zero magnetic field,~\cite{covington,biswas,dagan,sharoni,kohen}
other did not observe the splitting.~\cite{RPP,Ekin,Alff,wei,iguchi,Sawa1,Aubin}
Thus opinions are still divided among scientists on the BTRSS in high-$T_c$ superconductors.
If the BTRSS does not exist, we have to find another reasons for the peak
splitting observed in experiments.
In recent papers, we have showed that the interfacial randomness causes 
the split of the ZBCP in the zero magnetic field in both numerically~\cite{asano02-1} 
using the recursive Green function method~\cite{lee,asano01-1} and analytically~\cite{single}
using the single-site approximation.~\cite{asano96}
Our conclusion, however, contradicts to those of a number of theories
~\cite{barash2,golubov,poenicke,yamada,tanaka01b,luck}
based on the quasiclassical Green function 
method.~\cite{eilenberger,larkin,zaitsev,shelankov,bruder} 
The drastic suppression of the ZBCP by the interfacial randomness is 
the common conclusion of all the theories. The theories of the 
quasiclassical Green function method, however, concluded 
that the random potentials do not split the ZBCP.

\section{effects of magnetic field}
The TRS is also broken by applying external magnetic fields onto 
NS junctions. 
The resonance at $E=0$ is suppressed  because a quasiparticle acquires a 
Aharonov-Bohm like phase from magnetic fields.~\cite{asano00-1,asano00-2,asano00-3}
Actually it is pointed out that the ZBCP in NS junctions
splits into two peaks under the magnetic field.~\cite{fogelstrom,covington,YT021,YT022,YT023}
\begin{figure}[htbp]
\begin{center}
\includegraphics[width=8.0cm]{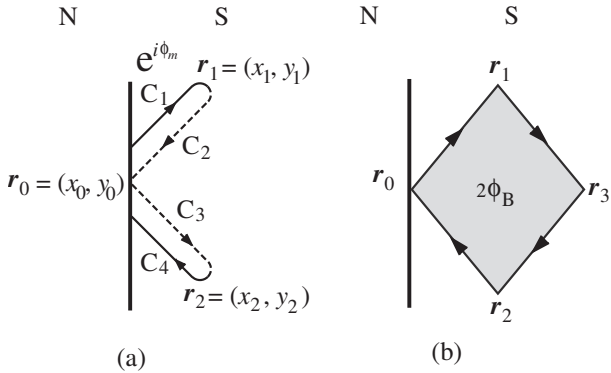}
\end{center}
\caption{
The motion of a quasiparticle near the interface is illustrated.
}
\label{promag}
\end{figure}
The reflection process in Fig.~\ref{promag}(a) corresponds 
to $A_s$ in Eq.~(\ref{a-def}).
 We consider uniform magnetic fields perpendicular to the $xy$ plane
(i.e., $B \hat{\boldsymbol{z}}$) and
assume the Landau gauge $\boldsymbol{A}_{ext} = Bx \hat{\boldsymbol{y}}$. 
Effects of magnetic fields are taken into account through the phase
of the wave function by using the gauge transformation. 
While traveling from $\boldsymbol{r}_0$ to 
$\boldsymbol{r}_1$, an electronlike quasiparticle acquires a phase $e^{i\phi_m}$
with
\begin{equation}
\phi_m=
\frac{e}{\hbar c}\int_{\boldsymbol{r}_0}^{\boldsymbol{r}_1} \!\!\!\! d\boldsymbol{l}
\cdot \boldsymbol{A}_{ext}(\boldsymbol{l})
=\frac{e}{\hbar c}\frac{B}{2}(x_1+x_0)(y_1-y_0).
\end{equation}
Since the magnetic field is sufficiently weak, the integration path 
can be replaced by a straight line between $\boldsymbol{r}_0$ and 
$\boldsymbol{r}_1$ which is denoted by $C_1$ in Fig.~\ref{promag}(a),
This approximation is justified when the radius of the cyclotron motion
of a quasiparticle, $2\mu_F/(k_F \hbar eB/mc)$, is much larger than $\xi_0$.
The condition is equivalent to the relation
$ \pi \Delta_0 \gg \hbar eB/mc$.
In high-$T_c$ materials, $\Delta_0 \sim 30-40$ meV, whereas $\hbar eB/mc$ 
is $10^{-1}$ meV for $B=1$ Tesla, where we use the bare mass of an electron.
The phase shift on the way from $\boldsymbol{r}_1$ to $\boldsymbol{r}_0$ ($C_2$)
in the hole branch is equal to $e^{i\phi_m}$.
This is because the direction of a quasiparticle's motion and 
the sign of the charge on $C_1$ are opposite to those on $C_2$ at the same time.
In the same way, we can show that the phase shifts on $C_3$ and $C_4$ in Fig.~\ref{promag} (a) 
are also $e^{i\phi_m}$.
Under the gauge transformation, 
the pair potential should be changed to 
\begin{equation}
\Delta(\boldsymbol{r},\boldsymbol{r}') 
\exp\left[ \frac{ie}{\hbar c}\left( \int^{\boldsymbol{r}}\!\!\!\!+ \int^{\boldsymbol{r}'}\right)  
d\boldsymbol{l}\cdot \boldsymbol{A}_{ext}(\boldsymbol{l})\right]. \label{gip}
\end{equation}
At $\boldsymbol{r}_1$, a phase factor 
\begin{equation}
\exp\left[ \frac{-i2e}{\hbar c}\int^{\boldsymbol{r}_1} \!\!\!\! 
d\boldsymbol{l}\cdot \boldsymbol{A}_{ext}(\boldsymbol{l})\right]
\end{equation}
is mulitiplied to the Andreev reflection coefficients, where 
$\boldsymbol{r}$ and $\boldsymbol{r}'$ in Eq.~(\ref{gip}) 
are set to be $\boldsymbol{r}_1$.
A phase factor
\begin{equation}
\exp\left[ \frac{i2e}{\hbar c}\int^{\boldsymbol{r}_2} \!\!\!\! 
d\boldsymbol{l}\cdot \boldsymbol{A}_{ext}(\boldsymbol{l})\right]
\end{equation}
is also multiplied to the Andreev reflection coefficients at $\boldsymbol{r}_2$.
The total phase shift by the magnetic field along $C_1 \sim C_4$ in 
Fig.~\ref{promag}(a) ($e^{i2\phi_B}$) is then given by
\begin{align}
\phi_B =& 2\phi_m + \frac{e}{\hbar c} 
\int^{\boldsymbol{r}_2}_{\boldsymbol{r}_1} \!\!\!\!
d\boldsymbol{l}\cdot \boldsymbol{A}_{ext}(\boldsymbol{l}),\\
=&-\frac{eB}{\hbar c}(y_1-y_0)(x_1-x_0)
= - \frac{B}{B_0}\frac{k_y}{k_x},\label{phib1}\\
B_0 =& \frac{\phi_0}{2\pi \xi_0^2},
\end{align}
where $\phi_0=2\pi \hbar c/e$.
On the way to Eq.~(\ref{phib1}), we use a relation 
%\begin{equation}
%\left( \begin{array}{c} x_1-x_0\\ y_1-y_0 \end{array}\right) // 
%\left( \begin{array}{c} k_x\\ k_y \end{array}\right),
%\end{equation}
$(x_1-x_0)/(y_1-y_0)=k_x/k_y$
and $x_1-x_0 \sim \xi_0$.
We note that $2\phi_B$ is the gauge invariant magnetic flux
passing through the gray area in Fig.~\ref{promag}(b), where 
$\boldsymbol{r}_3=(2x_1+x_0,y_0)$.
Thus $2\phi_B$ remains unchanged in another gauges such as 
$\boldsymbol{A}_{ext}=-By\hat{\boldsymbol{x}}$ and penetrating magnetic fields 
$\boldsymbol{A}_{ext}=B\lambda_0e^{-x/\lambda_0}\hat{\boldsymbol{y}}$
with $\lambda_0 \gg \xi_0$, where $\lambda_0$ is the penetration depth. 
In high-$T_c$ materials,
$\xi_0 \sim 2$nm and $\lambda_0 \sim 200$nm.
\begin{figure}[htbp]
\begin{center}
\includegraphics[width=8.0cm]{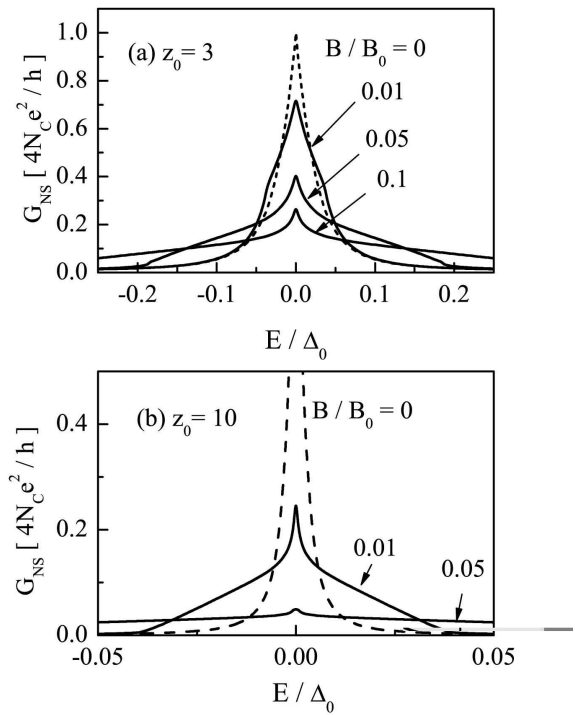}
\end{center}
\caption{
The conductance under magnetic fields for $d_{xy}$ symmetry, where
$z_0=3$ and 10 in (a) and (b), respectively. In high-$T_c$ material, 
$B_0$ is estimated to be 160 Tesla.
}
\label{fig:mag1}
\end{figure}

Effects of magnetic field can be taken into account in the present theory
by 
\begin{equation}
A_s \to A_s e^{2i\phi_B}, \label{asb}
\end{equation}
where $A_s$ is defined in Eq.~(\ref{a-def}). 
We show the conductance in $d_{xy}$ wave junctions calculated from Eqs.~(\ref{rhe1})-(\ref{btk1}) 
and (\ref{asb}) in Fig.~\ref{fig:mag1}, where $z_0=3$ and 
10 in (a) and (b), respectively. 
In high-$T_c$ superconductors, $B_0$ is about 160 Tesla.
The ZBCP decreases with increasing $B$ in both Figs.~\ref{fig:mag1} (a) and (b).
The degree of suppression due to magnetic fields depends on the transmission
probability of the junction. 
More drastic suppression can be seen in lower transparent junctions.
In Fig.~\ref{fig:mag1} (b), the ZBCP almost disappears for $B=0.05 B_0$. 
The ZBCP, however, remains one peak and does not split into two peaks even in the 
strong magnetic fields.
The results in Fig.~\ref{fig:mag1} are qualitatively well described by the 
analytical expression of the Andreev reflection probability for $E \ll \Delta_0$,
\begin{align}
|r^{he}|^2=& \frac{|t_N|^4}{|t_N|^4 + 4 |r_N|^2\sin^2(\theta+\phi_B)},\\
\simeq & \frac{|t_N|^4|\Delta|^2}{|t_N|^4|\Delta|^2 + 4 |r_N|^2 
( E+ |\Delta| \phi_B)^2}. \label{qca1}
\end{align}
We linearize the magnetic fields in $\sin(\theta+\phi_B)$ in Eq.~(\ref{qca1}).
Equation(\ref{qca1}) implies that the resonance energy may be shifted from $E=0$ by 
magnetic fields.
In contrast to the splitting of the ZBCP by the BTRSS in Sec.~IV, we do not 
find the peak splitting under magnetic fields in Fig.~\ref{fig:mag1}.
In the BTRSS, the shift of the resonance energy is caused by the $s$ wave component which has 
the resonance energy at $E=\Delta_0$.
On the other hand, any resonant states are not associated with magnetic fields.
Thus the magnetic fields only suppress the resonance of the ZES as shown in
Fig.~\ref{fig:mag1}.

\begin{figure}[htbp]
\begin{center}
\includegraphics[width=8.0cm]{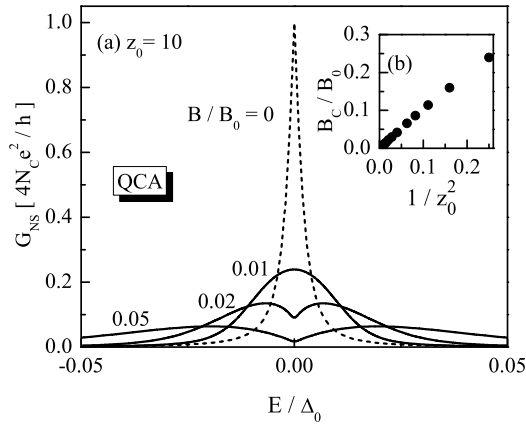}
\end{center}
\caption{
The conductance in the quasiclassical approximation 
is plotted for several magnetic fields in (a), where $z_0=10$.
In (b), threshold magnetic fields are shown as a function of $1/z_0^2$.
}
\label{fig:magqc}
\end{figure}

In a previous paper,~\cite{fogelstrom} however, the split of the ZBCP in magnetic fields 
was reported within the quasiclassical approximation (QCA).
The results in Eq.~(\ref{qca1}) are similar to that in the argument of the Dopplar 
shift in the QCA. 
The supercurrents flows along the interface shift
the energy of a quasiparticle as 
\begin{align}
E\to& E+\boldsymbol{v}_F \cdot \boldsymbol{p}_s, \\
\boldsymbol{p}_s=& -\frac{e \boldsymbol{A}}{c} = 
\frac{eB \lambda_0 }{c} e^{-x/\lambda_0} \hat{\boldsymbol{y}},\label{psqca}
\end{align}
where $\boldsymbol{p}_s$ is the condensate momentum at the interface.
In Eq.~(\ref{psqca}), $d$ wave character of the supercurrent is not considered.
The corresponding approximation in the present theory is replacing
$E+ |\Delta| \phi_B$ by $E+ \Delta_0 \phi_B$ in Eq.~(\ref{qca1})
and we find
\begin{equation}
|r^{he}|^2_{\textrm{QCA}}= \frac{|t_N|^4|\Delta|^2}{|t_N|^4|\Delta|^2 + 4 |r_N|^2 
( E+ \Delta_0 \phi_B)^2}. \label{qca2}
\end{equation}
In Fig.~\ref{fig:magqc} (a), we show the conductance calculated from Eq.~(\ref{qca2})
for $z_0=10$.
In contrast to Fig.~\ref{fig:mag1} (b), we find split of the ZBCP when magnetic fields
are larger than the threshold magnetic field, $B_c$. 
The threshold depends on $z_0$ as shown in Fig.~\ref{fig:magqc} (b), where $B_c$ is 
plotted as a function of $1/z_0^2$ which is proportional to the normal transmission 
probability of junctions.
The threshold increases with increasing the transmission probability of junctions. 
This has been pointed out in the
conductance calculated on the lattice model by using the QCA.~\cite{YT022} 
In the lattice model, it was also shown that 
$B_c$ decreases with the increase of the doping rate. The Fermi energy is 
a decreasing function of the doping rate. Therefore the transmission probability
of junctions decreases with increasing the doping rate.

Although Eqs.~(\ref{qca1}) and (\ref{qca2}) are similar to each other, the
response of the ZBCP to magnetic fields are qualitatively different.
To make clear if a magnetic field splits the ZBCP or not, we need some numerical 
simulations, where effects of magnetic field are taken into account accurately.
In experiments, some papers show the split of the ZBCP in magnetic 
fields.~\cite{covington,dagan}
On the other hand, several papers report no splitting of the 
ZBCP.~\cite{Ekin,qazilbash,allf,sawa2} 
A microscopic scattering theory indicates that the sensitivity of the ZBCP
to magnetic fields depends on the degree of potential disorder near 
the NS interface.~\cite{single}

Finally we briefly discuss an important difference of 
the conductance in the present theory and that in the QCA.
The phenomenological theory reaches at Eq.~(\ref{qca2}) which is almost the same 
as the conductance expression in the QCA.~\cite{fogelstrom}
The two theories, however,  still lack a quantitative agreement of
the threshold magnetic field. 
The normalization for the penetrating magnetic fields in the QCA  
($B_0^{QCA} = \phi_0/(2\pi \xi_0\lambda_0)$) is about 1.6 Tesla with 
$\lambda_0 \sim 100 \xi_0$.~\cite{fogelstrom,YT022} 
In Eq.~(\ref{psqca}), $\boldsymbol{p}_s$ in the QCA is originally given by the 
vector potential which is not an observable value. Thus the QCA does not satisfy
the gauge invariance. 
In the present theory, on the other hand, we consider uniform magnetic field and the 
normalization of magnetic fields ($B_0$) is about 160 Tesla. 
This value remains unchanged even if we consider penetrating magnetic field as
$Be^{-x/\lambda_0} \hat{\boldsymbol{z}}$ with $\lambda_0 \gg \xi_0$.
For example in Fig.~(\ref{fig:magqc}) (b), we find that $B_c$ is about $0.1B_0$ at $z_0=3$. 
Therefore $B_c$ is estimated to be 16 Tesla in the present theory. 
The same results are interpreted as $B_c$=0.16 Tesla if we use $B_0^{QCA}$ in the QCA.
The threshold magnetic field in the QCA is estimated to be much smaller than that 
in the present theory. This disagreement may be important because the maximum value of
magnetic fields in experiments is about 10 Tesla.~\cite{Ekin,qazilbash,allf,sawa2}

\section{conclusion}
We have presented a phenomenological theory of the
Andreev reflection to make clear reasons for the appearance of the zero-bias conductance 
peak (ZBCP) in normal-metal / unconventional superconductor junctions.
The phenomenological theory reveals that the zero-energy state (ZES) is a consequence of
the constructive interference effect of a quasiparticle. 
The expression of the Andreev reflection probability enables us to understand an 
importance of the unconventional pairing symmetry for the formation of the ZES.
The phenomenological theory is applied to superconductors 
with a broken time-reversal symmetry state (BTRSS) and junctions under magnetic fields.
The split of the ZBCP in $s+id_{xy}$ wave superconductors is understood in terms of 
the shift of the resonance energy by the $s$ wave component.  
The Aharonov-Bohm like phase received from magnetic fields suppresses the degree of resonance 
of the ZES, which explains the suppression of the ZBCP in magnetic fields.

\appendix

\section{Andreev reflection by spin-triplet superconductors}
In the text, we consider two-dimensional spin-singlet superconductors and $\delta$-function type
potential barrier for simplicity.
Here we generalize the phenomenological theory to spin-triplet superconductors in three-dimension.
The pair potential in superconductors is given by
\begin{align}
\hat{\Delta}(\boldsymbol{k})=& \begin{cases} i 
\boldsymbol{d}(\boldsymbol{k})\cdot \hat{\boldsymbol{\sigma}}\hat{\sigma}_2 & 
\text{: triplet}, \\
i d(\boldsymbol{k}) \hat{\sigma}_2 & \text{: singlet},   
\end{cases}
\end{align}
where $\hat{\sigma}_j$ for $j=1$, 2 and 3 are Pauli matrices representing the 
spin degree of freedom.
We assume that the current is in the $x$ direction and consider a potential 
barrier
\begin{equation}
V(\boldsymbol{r})= V_0 \left[ \Theta(x) - \Theta(x-L)\right], \label{v01}\\
\end{equation}
where $L$ is the thickness of the insulating layer.
The Andreev reflection coefficients in the absence of the insulator
are calculated analytically 
\begin{align}
\hat{r}^{he}_0 =& - e^{-i\varphi_s} \hat{\Delta}_{(+)}^\dagger \hat{R}_{(+)}, \\
\hat{r}^{eh}_0 =& - e^{i\varphi_s} \hat{R}_{(-)}\hat{\Delta}_{(-)}, \\
\hat{\Delta}_{(\pm)} =& i \boldsymbol{d}_\pm \cdot \hat{\boldsymbol{\sigma}} \hat{\sigma}_2,\\
\hat{R}_{(\pm)} =& \frac{1}{2|\boldsymbol{q}_\pm|}\sum_{l=1}^2
\left[ \frac{K_{l,\pm}}{\Delta_{l,\pm}^2} \hat{P}_{l,\pm}\right],
\end{align}
\begin{align}
\Delta_{l,\pm}=&\sqrt{ |\boldsymbol{d}_\pm|^2 -(-1)^l |\boldsymbol{q}_\pm|},\label{delnu}\\
K_{l,\pm}=& \sqrt{E^2-\Delta_{l,\pm}^2}-E,\\
\hat{P}_{l,\pm} =& |\boldsymbol{q}_\pm| \hat{\sigma}_0 -(-1)^l \boldsymbol{q}_\pm \cdot
\hat{\boldsymbol{\sigma}},\\
\boldsymbol{q}_\pm =& i \boldsymbol{d}_\pm \times \boldsymbol{d}_\pm^\ast,\\
\boldsymbol{d}_\pm =& \boldsymbol{d}(\pm k_x,k_y,k_z),
\end{align}
where $\varphi_s$ is a macroscopic phase of superconductor,
$l(=1$ or 2) indicates the two spin branches of Cooper pairs and $\hat{\sigma}_0$
is the $2\times 2$ unit matrix.
The normal transmission and the normal reflection coefficients of the insulator
are calculated as
\begin{align}
\hat{t}_N =& \frac{- 2i \bar{k}_x \bar{p}_x e^{-ik_xL}}{z_1^\ast}\hat{\sigma}_0,\\
\hat{r}_N =& \frac{-z_0}{z_1^\ast}\hat{\sigma}_0,\\
z_0 =& \frac{V_0}{\mu_F}\sinh(p_xL),\\
z_1=& (\bar{p}_x^2-\bar{k}_x^2) \sinh(p_xL) + 2i\bar{k}_x\bar{p}_x \cosh(p_xL),
\end{align}
where 
$p_x = \sqrt{(V_0/\mu_F) - (k_x/k_F)^2}$ is the wave number at the insulator
and $\bar{p}_x=p_x/k_F$.

The argument in Sec.~II leads to the exact expression of the Andreev and the normal 
reflection coefficients~\cite{asano03-3} which are given by 
\begin{align}
\hat{r}^{ee} =& - z_0z_1 \left[ \hat{\sigma}_0 - \hat{W}\right] 
\left[ |z_1|^2\hat{\sigma}_0 - z_0^2\hat{W}\right]^{-1}, \label{tree}\\
\hat{r}^{he} =& - e^{-i\varphi_s} 4\bar{k}_x^2\bar{p}_x^2  
\hat{\Delta}_{(+)}^\dagger \hat{R}_{(+)}
\left[ |z_1|^2\hat{\sigma}_0 - z_0^2\hat{W}\right]^{-1},\\ 
\hat{W} =& \hat{R}_{(-)} \hat{\Delta}_{(-)} \hat{\Delta}_{(+)}^\dagger \hat{R}_{(+)}.\label{defw}
\end{align}
The results of unitary states including the spin-singlet states can be obtained when
we use following relations
\begin{align}
\hat{R}_{(\pm)} =& \frac{\sqrt{E^2-|D_\pm|^2}-E}{|D_\pm|^2}\hat{\sigma}_0,\\
|D_\pm| =& \left\{ \begin{array}{cc} |d_\pm| & \textrm{: singlet} \\
                                |\boldsymbol{d}_\pm| & \textrm{: triplet},
                   \end{array} \right.       
\end{align}
in Eqs.~(\ref{tree})-(\ref{defw}). 
The differential conductance is given by
\begin{align}
G_{NS}=&\frac{e^2}{h}\left.\sum_{k_y, k_z} \textrm{Tr} \left[ 
\hat{\sigma}_0 - \hat{r}^{ee} (\hat{r}^{ee})^\dagger + \hat{r}^{he} 
(\hat{r}^{he})^\dagger \right]
\right|_{E=eV_{bias}}.
\end{align}
A relation $\boldsymbol{d}_- =- \boldsymbol{d}_+$
represents the condition for the perfect formation of the ZES. 
Actually when $\boldsymbol{d}_+ =\boldsymbol{d}=\nu \boldsymbol{d}_- $ with $\nu =\pm 1$,
the Andreev reflection probability becomes
\begin{equation}
\textrm{Tr} \left[\hat{r}^{he} (\hat{r}^{he})^\dagger \right]=
\sum_{l=1}^{2} \left|\frac{4\bar{k}_x^2\bar{p}_x^2 \Delta_l K_l}
{4\bar{k}_x^2\bar{p}_x^2 \Delta_l^2 + z_0^2 (\Delta_l^2 - \nu K_l^2)}\right|^2,
\end{equation}
where $K_l=K_{l,+}=K_{l,-}$ and $\Delta_l =\Delta_{l,+}=\Delta_{l,-}$.
 
In the limit of $E\to 0$ and $z_0 \gg 1$, we find 
\begin{equation}
\textrm{Tr} \left[\hat{r}^{he} (\hat{r}^{he})^\dagger \right]= \left\{ \begin{array}{ccc} 
2 \left(\frac{ 4\bar{k}_x^2\bar{p}_x^2}{2 z_0^2}\right)^2 & :& \nu=1 \\
 & & \\
 2 & :& \nu = -1, \end{array}\right.
\end{equation}
where spin degree of freedom give rise to a factor 2.
Thus the zero-bias conductance is independent of the transmission probability
of junctions when 
$\boldsymbol{d}_- =- \boldsymbol{d}_+$ is satisfied.

In spin-singlet superconductors, we show that the internal phase of a Cooper pair is 
responsible for the ZES. 
In spin-triplet superconductors, the internal spin degree of freedom of a Cooper pair
has another possibilities for the formation of some resonant states in subgap energies.

\end{document}